\documentclass[a4paper,twocolumn,showpacs]{revtex4-1}

\usepackage{latexsym}
\usepackage[dvipdf]{graphicx}
\usepackage{mathrsfs}
\usepackage{color}

\newcommand{\ket}[1]{\left| #1 \right\rangle} % for Dirac bras
\newcommand{\bra}[1]{\left\langle #1 \right|} % for Dirac kets

\newcommand{\ketbra}[2]{\ket{#1}\!\bra{#2}}
\newcommand{\im}{{\mathrm{i}}}
\newcommand{\re}{{\mathrm{Re}}}

\begin{document}

\title[Control protocol of finite quantum systems]
{Control protocol of finite dimensional quantum systems using
alternating square pulse}

\author{Jianju Tang and H. C. Fu \footnote{E-mail: hcfu@szu.edu.cn, corresponding author.} }

\address{School of Physical Sciences and Technology, Shenzhen University,\\
Shenzhen 518060, P. R. China}

\begin{abstract}
Control protocol to drive finite dimensional quantum systems to an arbitrary target state using
square pulses is proposed explicitly. It is a multi-cycle control process and in each cycle we apply
square pulses to cause single or a few transitions between energy levels. Systems with equal energy
gaps except the first one, four dimensional system with equal first and third energy gaps and different
second energy gap, and systems with all equal energy gaps of dimension three, are investigated
in detail. The control parameters, the interaction time between systems and control fields and free
evolution times between cycles, are connected with the probability amplitudes of target states via
trigonometric functions and are determined analytically.
\end{abstract}

\maketitle

\section{Introduction}

Quantum control is a coherent or incoherent process to
steer a quantum system to a given target state \cite{1}. It is
significant in many fields of quantum physics, especially
in the quantum computation and quantum information
processing \cite{ref1}. Various notations in classical control
theory were generalized to the quantum control, such as open
and closed control, optimal control \cite{optimal}, controllability
\cite{controllability,fu1,fu2,
graph}, feedback control \cite{feedback} and so on.
For incoherent control
the quantum system is controlled by its interaction with a
quantum accessor rather than classical fields \cite{indirect1,indirect2,romano,penchen}. Typically
one first models the controlled system and examine
its controllability which is relevant to the system Hamiltonian
and interaction Hamiltonian with control fields.
Then one needs to design classical fields to interact with
the controlled system to steer the system to the given target
state, which is referred to as the control protocol and
is the issue we would like to address in this paper. Some
approaches, such as using the Cartan decomposition of
Lie groups [13], were proposed on this issues.

In a previous paper [14], we proposed an explicit control
protocol of finite dimensional quantum system using
time-dependent cosine classical field, where we have to
use the rotating wave approximation to drop the high
oscillating terms in the interaction Hamiltonian. In this
work, we will use the alternating square pulses with positive and negative
parts in each period, to control the finite quantum
systems. Advantage of using square pulses is that we can
find an invariant space in which the interaction Hamiltonian
and time evolution operator can be explicitly treated
without using rotating wave approximation. On the other
hand, except the system with equal energy gaps except
the first we considered before, we mainly consider the
systems with multiple transitions in a cycle, especially
the one with all equal adjacent energy gaps of dimension
three. The relationship between probability amplitudes
of target states and control parameters, the number of
pulses and the difference between positive and negative
amplitudes of the pulses, is analytically established.

This paper is organized as follows. In Sec. II, we formulate
the controlled system and control protocol, and
derive the time evolution operators. We present control protocols
of systems with equal energy gaps except the first one,
four dimensional system with equal first and third energy
gaps and different second energy gap, and system with
equal energy gaps of dimension three in sections III, IV,
and V, respectively. We draw our conclusion in Sec.VI.

Throughout this paper we use $\im = \sqrt{-1}$ and $\hbar=1$.

\section{Control systems and time evolution operators}

\subsection{Control Systems and protocol}

Consider an $N$-dimensional non-degenerate quantum system with eigenenergy $E_n$ and
corresponding eigenstate $\ket{n}$, described by the Hamiltonian
\begin{equation}\label{eq2.1}
    H_0=\sum^N_{n=1} E_n \ket{n}\!\bra{n}.
\end{equation}
Our aim is to develop control protocols to drive the system to an arbitrary target states
from an initial state, or in other words,
to design classical fields to
interact with the system such that the system is driven to a required
target state within finite time.
For this purpose, we should keep in mind that
\begin{itemize}
\item  As there are $2(N-1)$ independent real parameters
in the target states ($N-1$ real probability amplitudes and $N-1$
relative phases), we need to supply $2(N-1)$ real control parameters in the
control protocol;

\item  One needs to establish relationship between
control parameters and probability amplitudes of the target states
such that one can design the control fields and its coupling with the
controlled systems;

\item In each cycle the control operation should be easily complemented in
the laboratory.
\end{itemize}

\begin{figure}[tbp]
\centerline{\includegraphics[scale=0.6,width=7cm]{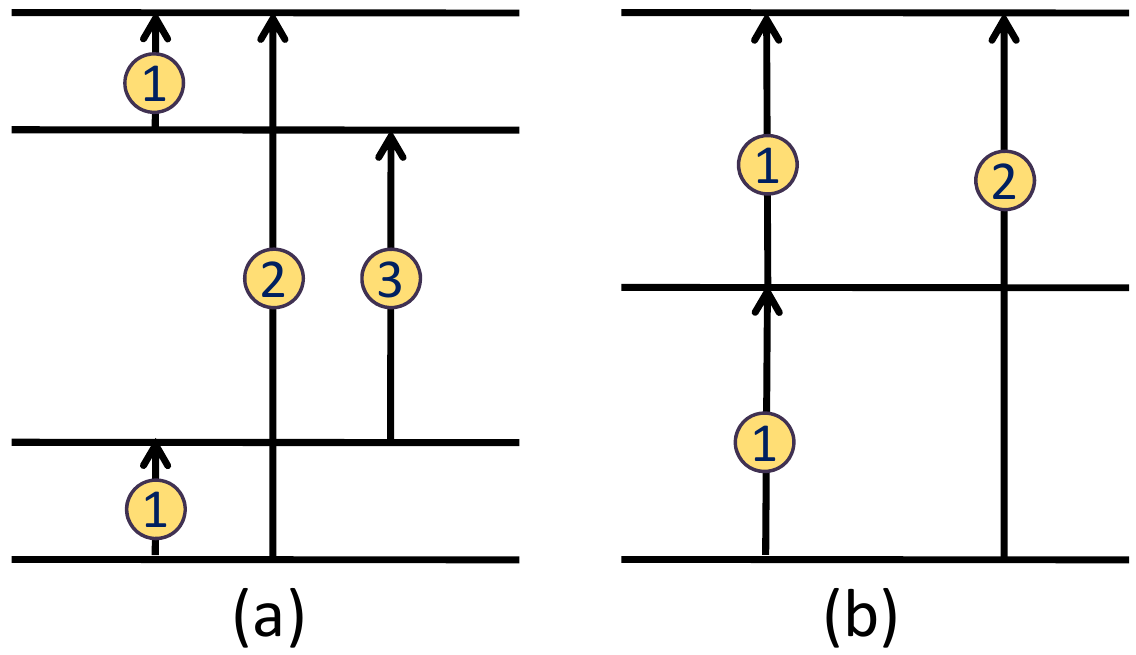}}
\caption{
Four dimensional system with $\mu_1 = \mu_3 \neq \mu_2$ (System II), and
3-dimensional system with equal energy gaps (System III). Here the arrows
represent transitions caused by control field and the numbers in circles are
labels of cycle. }
\label{tang2-4d-system}
\end{figure}

The simplest protocol is that one can control transition only
between two energy levels in each cycle, for example, the system
(System I) with all equal energy gaps except the first one
\begin{equation}
\mu_1 \neq \mu_2=\mu_3=\cdots = \mu_{N-1},
\end{equation}
where $\mu_i = E_{i+1}-E_i$ is the energy gap,
and the system with all distinct adjacent energy gaps $\mu_i\neq \mu_j$ ($i\neq j$),
considered in Ref.~\cite{tang1}.
However, this is not always possible,
for example, the 4-dimensional system with $\mu_1 = \mu_3 \neq \mu_2$ as
shown in Fig.\ref{tang2-4d-system}(a) (System II). In this case, the control field with frequency $\mu_1$
cause transition form $\ket{1}$ to $\ket{2}$ and from $\ket{3}$ to $\ket{4}$ simultaneously and
the interaction Hamiltonian is written as
\begin{equation}
H= H_0 + f(t)\left(\ketbra{1}{2}+\ketbra{2}{1}+\ketbra{3}{4}+\ketbra{4}{3}\right).
\end{equation}
Fortunately, its time evolution operator can be factorized. In particular, for some
special initial state such as $\ket{1}$, if we arrange this control process as the
first cycle, the states $\ket{3}$ and $\ket{4}$ keep unchanged and mathematically
equivalent to only transition between $\ket{1}$ and $\ket{2}$ occurs. We will
investigate control protocol of this system in Sec.\ref{section-4dimensional}.

However, for systems with all equal adjacent energy gaps, the time evolution operator
is difficult to treat explicitly. For 3-dimensional system as shown in Fig.1(b), the
Hamiltonian in the cycle 1 is
\begin{equation}
H= H_0 + f(t)\left[d_1(\ketbra{1}{2}+\ketbra{2}{1})+d_2(\ketbra{2}{3}+\ketbra{3}{2})\right].
\end{equation}
Although the time evolution operator can be generally factorized in the form
\begin{equation}
U(t) = \prod_{i=1}^{\dim L} \exp\left[\alpha_i (t) X_i\right],
\end{equation}
where $X_i$'s are a basis of Lie algebra $L=\mbox{u}(3)$ (or su(3) if $H_0$ is traceless),
according to Wei-Norman theorem \cite{wei}, the functions $\{\alpha_{i}(t)\}$ satisfy a set of non-linear
equations and are difficult to find explicitly. In Sec.V, we present an explicit approach to
obtain the time evolution operator.

\subsection{Pules and time evolution operator}\label{sec-pulse-evolution-operator}

In this paper, we shall use square pulses to control the quantum systems.
We shall see that one does not need to use the rotating wave approximation
in the derivation of time evolution operators as we did in \cite{tang1}.

\begin{figure}[tbp]
\centerline{\includegraphics[scale=0.6,width=8cm]{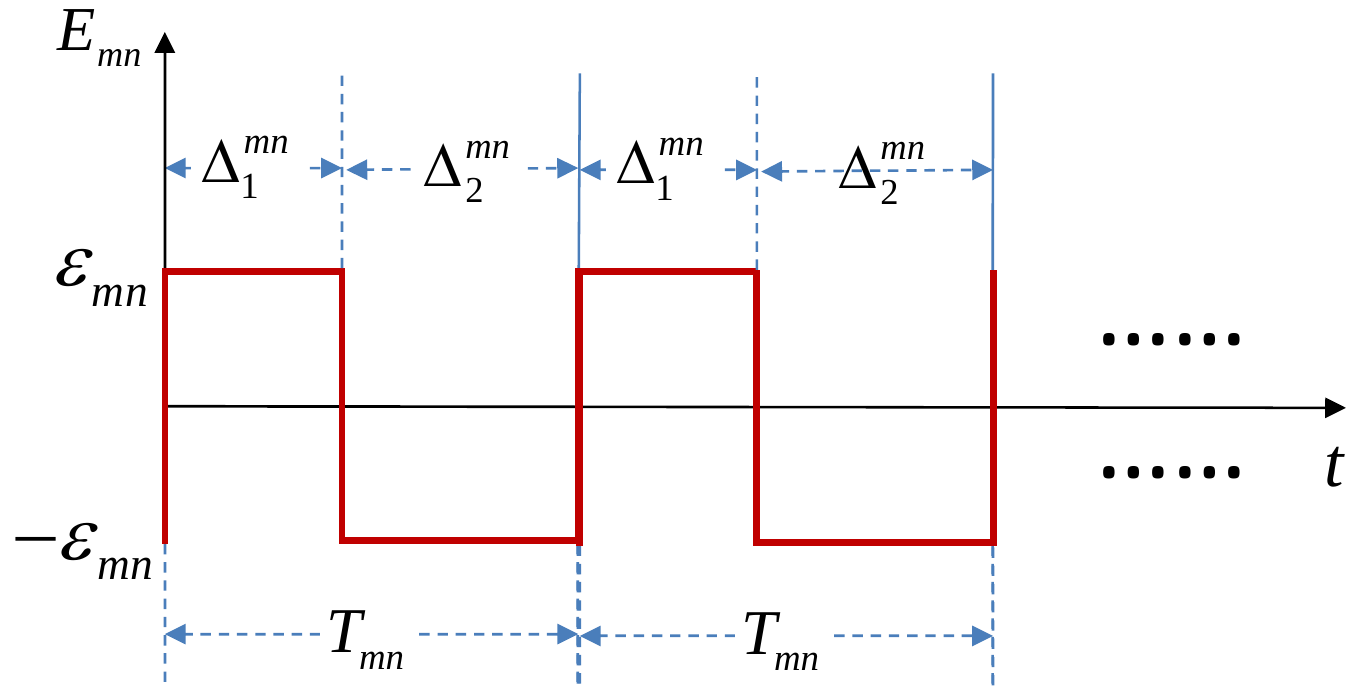}}
\caption{Control pulses. }
\label{tang2-pules}
\end{figure}

Consider a sequence of pulses interacting only with two levels $\ket{m}$
and $\ket{n}$ ($m>n$) resonantly, which will be used
in the rest of this paper. Each pulse can
be written as
\begin{equation}
E_{mn}(t) = \left\{
\begin{array}{ll}
{\cal E}_{mn}, &   0 \leq t \leq  \Delta^{mn}_1; \\
{\cal -E}_{mn}, &  \Delta_1^{mn} \leq t \leq \Delta^{mn}_1+\Delta^{mn}_2,
\end{array}
\right.
\end{equation}
as shown in Fig.~(\ref{tang2-pules}). Here ${\cal E}_{mn}$ is constant electric field
and $T_{mn}=\Delta^{mn}_1+\Delta^{mn}_2$ is its period.
Condition of resonant interaction requires that
\begin{equation}
\omega_{mn}=2\pi/T_{mn} = E_{m}-E_n.
\end{equation}

Corresponding to $E_{mn}(t)=\pm {\cal E}_{mn}$, the total
Hamiltonian of the system and control pulse is in the form
\begin{eqnarray}
   H^{(mn)}_\pm = H_0 \pm d_{mn} \sigma^{mn}_x,  \label{Hamiltonian-pules}
\end{eqnarray}
where $d_m = {\cal E}_m g$,  $g$ is the coupling strength, and
\begin{equation}
\sigma^{mn}_x \equiv \ketbra{m}{n}+\ketbra{n}{m}.
\end{equation}

As the Hamiltonian (\ref{Hamiltonian-pules}) is time independent, the corresponding time evolution
operator can be written as $U^{mn}_\pm(t) = \exp\left(\im H_\pm^{(mn)}t\right)$.
To expand it to a simple form, we first rewrite the Hamiltonian as
%\begin{eqnarray}\label{free-ham}
%H_0
%   &=& \frac{1}{2}\omega_{mn}(\ketbra{n}{n}-\ketbra{m}{m} )+\theta_{mn} I       \nonumber\\
%   & & +\sum^N_ {k=1\atop k \neq m,n}\left(E_k-\theta_{mn}\right)\ketbra{k}{k},
%\end{eqnarray}
\begin{eqnarray}\label{new-form-ham}
H^{(mn)}_\pm=H^{(mn)}_0 + H^{(mn)}_{c\pm},
\end{eqnarray}
where
\begin{eqnarray}
&& H^{(mn)}_0= \theta_{mn} I
   +\sum^N_ {k=1\atop k\neq m,n}\left(E_k-\theta_{mn} \right)\ketbra{k}{k}, \nonumber\\
%   &=&\frac{1}{2}(E_1+E_ {m+1})I+\sum^N_ {n=2\atop n\neq m+1}B_n\ketbra{n}{n}\\
&& H^{(mn)}_{c\pm} = \frac{1}{2}\omega_{mn}(\ketbra{n}{n}-\ketbra{m}{m} )
    \pm d_{mn} \sigma_x^{mn},\nonumber \\
&& \theta_{mn}=\frac{1}{2}(E_n+E_{m}),
\end{eqnarray}
and $I$ is the identity operator.
It is obvious that
%with $\omega_m=E_m-E_{m+1}$.
%Correspondingly, the total Hamiltonian can be rewritten as
\begin{equation}
\left[H^{(mn)}_{c\pm}, H^{(mn)}_0\right]=0. \label{hh-commute}
\end{equation}

Noticing that
\begin{equation}
H_0^{(mn)}\ket{k} = \left\{
\begin{array}{ll}
\theta_{mn} \ket{k}, &   k=m,n, \\
E_k \ket{k}, & k \neq m,n,
\end{array}
\right.
\end{equation}
the subspace ${\cal H}_{mn}$ spanned by $\{\ket{n}, \ket{m}\}$ is invariant
under the action of $H_0^{(mn)}$ and thus $H^{(mn)}_{\pm}$.

It is easy to verify that
\begin{eqnarray}
   && \left(H^{(mn)}_{c\pm}\right)^{2k}=( \Omega_{mn})^{2k} I_{mn},
      \quad    (k\geq 1), \nonumber \\
   && \left(H^{(mn)}_{c\pm}\right)^{2k+1}=( \Omega_{mn})^{2k}H^{(mn)}_{c\pm},
      \quad   (k\geq 0),
      \label{simple-relation-hc}
\end{eqnarray}
where
\begin{equation}
     \Omega_{mn}=\sqrt{\frac{1}{4}\omega_{mn}^2+d_{mn}^2},
\end{equation}
and $I_{mn} \equiv \ketbra{n}{n}+\ketbra{m}{m}$ is the identity operator
on the invariant subspace ${\cal H}_{mn}$.
Using relations (\ref{hh-commute}) and (\ref{simple-relation-hc}),
we find the time evolution operator
\begin{eqnarray}\label{time-evo-operator}
   && U^{(mn)}_\pm (t)=e^{ -\im \left(H^{(mn)}_0+H^{(mn)}_{c\pm}\right)t}
      =e^{ -\im H^{(mn)}_0t}e^{-\im H^{(mn)}_{c\pm}t}   \nonumber\\
   && \quad = \left\{I+\left[\cos(\Omega_{mn}t)-1\right] I_{mn}
      -\frac{\im \sin(\Omega_{mn}t)}{\Omega_{mn}}H^{({mn})}_{c\pm}\right\} \nonumber\\
   && \qquad  \times e^{ -\im H^{({mn})}_0t} .
\end{eqnarray}
In the invariant subspace ${\cal H}_{mn}$, we have
\begin{eqnarray}\label{U-effect-on-state1x}
   & & U^{({mn})}_\pm(t)\ket{n} =
       \left\{ \cos(\Omega_{mn}t)\ket{n}-\frac{\im \sin(\Omega_mt)}{\Omega_m} \right. \nonumber \\
   & & \quad  \times \left.
       \left(\frac{\omega_{mn}}{2}\ket{n}\pm d_{mn}\ket{m}\right)\right\}
       e^{-\im \theta_{mn}t}.
\end{eqnarray}
For simplicity, we suppose that the control field is strong, namely,
$d_{mn} \gg \omega_{mn}$. In this case, $\omega_{mn}/\Omega_{mn}\rightarrow 0$
and $d_{mn}/\Omega_{mn}\rightarrow 1$,
(\ref{U-effect-on-state1x}) is simplified as
\begin{eqnarray}\label{U-effect-on-state1-a}
   U^{({mn})}_\pm(t)\ket{n}
   &=& e^{-\im \theta_{mn}t}\left[\cos(\Omega_{mn}t)\ket{n} \right. \nonumber\\
   &&  \left.\mp \im \sin(\Omega_{mn}t)\ket{m}\right].
\end{eqnarray}
Similarly, we have
\begin{eqnarray}\label{1-m+1-change}
U^{({mn})}_{\pm}(t)\ket{m}
    &=& e^{-\im \theta_{mn}t}\left(\cos(\Omega_{mn} t)\ket{m} \right. \nonumber \\
    & & \left.\mp \im \sin(\Omega_{mn} t)\ket{n}\right).
\end{eqnarray}

In a period $T_{mn}$ of control field, the time evolution operator is
$U^{({mn})}(T_{mn})\equiv U^{({mn})}_- (\Delta^{mn}_2)U^{({mn})}_+(\Delta^{mn}_1)$, whose action
on the states $\ket{n}$ and $\ket{m}$ is obtained as
\begin{eqnarray}
&& U^{({mn})}(T_{mn}) \ket{n}
= e^{-\im\theta_{mn} T_{mn}}\left[\cos\left(\Omega_{mn} \Delta_{mn} \right)\ket{n} \right.\nonumber \\
&& \qquad \left. + \im \sin\left(\Omega_{mn} \Delta_{mn}\right)\ket{m}\right], \nonumber \\
&& U^{(mn)}(T_m) \ket{m}
= e^{-\im \theta_{mn} T_{mn}}\left\{\cos\left[\Omega_{mn} \Delta_{mn}\right]\ket{m} \right. \nonumber \\
&& \qquad   \left.  + \im \sin\left[\Omega_{mn} \Delta_{mn}\right]\ket{n}\right\},
\label{evolution-operator-a-period}
\end{eqnarray}
where $\Delta_{mn} \equiv \Delta^{mn}_2-\Delta^{mn}_1$.

Suppose that the system interacts with control field for $l_{mn}$ pulses (or time period
$\tau_{mn}=T_{mn}l_{mn}$). Then the time evolution operator is
\begin{equation}
    U^{({mn})}(\tau_{mn}) = U^{({mn})}(l_{mn} T_{mn}) = \left(U^{({mn})}(T_{mn})\right)^{l_{mn}},
\end{equation}
which can be obtained by replacing $\theta_{mn}$ by $l_{mn}\Delta_{mn}$
and $\Delta_{mn}$ by $l_{mn}\Delta_{mn}$ from $U^{({mn})}(T_{mn})$. Notice
that the real probability amplitude is determined by
$\Omega_{mn}l_{mn}\Delta_{mn}$ and one can adjust the parameter $\Delta_{mn}$
to insure that $l_{mn}$ is a positive integer.

Before closing this section, let us see the simplest case, the 2-level system.
It is obvious that, if the system is initially prepared on $\ket{1}$, it
is driven to the state
\begin{eqnarray}
&& \ket{\psi} = e^{- \im \theta_{12}l_{12}T_{12}} \nonumber \\
&& \quad \times \left[\cos(\Omega_{12}\Delta_{12}l_{12})\ket{1}
+ \im \sin(\Omega_{12}\Delta_{12}l_{12})\ket{2}\right]
\end{eqnarray}
after interesting with the control field for $l_{12}$ pulses.
We see that there is only a fixed relative phase $i=e^{i\pi/2}$, different from
the protocol using harmonic field given in \cite{tang1} in which there is a
relative phase $e^{-i(E_2-E_1)\tau_1}$ ($\tau_1$ is the interaction time) and it
is tunable through amplitude of control field. In the present case, we can allow
the system for a free evolution for time period $\tau_1^\prime$
\[
\ket{\psi^\prime} = \cos(\Omega_{12}\Delta_{12}l_{12})\ket{1}
+ \im e^{-\im\omega_{12}\tau_1^\prime}\sin(\Omega_{12}\Delta_{12}l_{12})\ket{2},
\]
up to a common phase $\delta = e^{-i\theta_{12}l_1T_{12}}e^{-iE_1\tau_1^\prime}$.
By carefully choosing control parameters
$\Omega_{12}\approx {\cal E}_{12}g, \Delta_{12}, l_{12}$ and $\tau_1^\prime$, we can
achieve any target state.

\section{Control protocol of system I}

In this section we investigate the control protocol of System I.
The control process includes $N-1$ cycles and in the $m$-th cycle, we first apply
control field with frequency $\omega_m = E_{m+1}-E_1$ to control transition
between levels $\ket{1}$ and $\ket{m+1}$ for a time period $\tau_m$ and then allow
the system to evolve for time period $\tau_m^\prime$. For convenience, we change
the label $(m+1\ 1) \to (m)$, for example, the time evolution operator is
relabeled as $U^{(m)}\equiv U^{(m+1\ 1)}$.
We will derive the explicit expression of target states and relationship
between the control parameters $\{\tau_m, \tau_m^\prime\}$
and probability amplitudes of target state.

\subsection{Cycle 1}

Suppose that the system is initially prepared on the state $\ket{1}$. Then it is easy to
find the state after interaction with control field for time $\tau_1$
%\[
%   \ket{\psi}^{(1)}
%   = e^{-\mathrm{i}\theta_1 l_1 T_1}\left[ \cos(\Omega_1 l_1\Delta_1)\ket{1}
%       + \im \sin(\Omega_1 l_1\Delta_1)\ket{2}\right].
%\]
and free evolution for $\tau'_1$
\begin{eqnarray} \label{1-cyc-free-evol}
\ket{\psi'}^{(1)} = % e^{-\mathrm{i}H_0\tau'_1}\ket{\psi}^{(1)}
      a^1_1\ket{1}+a^1_2\ket{2},
\end{eqnarray}
where
\begin{eqnarray}\label{1-cycle-coefs}
   && a^1_1=e^{-\mathrm{i}(\theta_1l_1 T_1+E_1\tau'_1)}\cos(\Omega_1 l_1\Delta_1),        \nonumber   \\
   && a^1_2=e^{-\mathrm{i}(\theta_1l_1 T_1+E_2\tau'_1)}\mathrm{i}\sin(\Omega_1 l_1\Delta_1).
\end{eqnarray}

\subsection{Recursion relations of amplitudes from $(m-1)$-th to $m$-th cycles}

To obtain the explicit expression of the target state, we need recursion relations of
probability amplitudes of states of two adjacent cycles. Suppose that, after $(m-1)$-th cycle, the
state is of the form
\begin{equation}\label{m-1-cyc-finalSta}
  \ket{\psi^\prime}^{(m-1)}=\sum^m_{k=1}a^{m-1}_k\ket{k}.
\end{equation}
Then the state of system after interacting with $l_m$ pulses and free evolution for a
time period $\tau_m^\prime$ can be obtained as
\begin{eqnarray}\label{l-pulses-exerted}
\ket{\psi^\prime}^{(m)}
   &=& e^{-\im H_0\tau_m^\prime} U^{(m)}(l_m T_m) \ket{\psi}^{(m-1)}     \nonumber\\
   &=& e^{-\mathrm{i}(\theta_ml_m T_m+E_1\tau_m^\prime)}\cos(\Omega_m l_m\Delta_m)a^{m-1}_1\ket{1}  \nonumber \\
   & & +\sum^m_{k=2}e^{-\mathrm{i}(E_k l_m T_m+E_k\tau_m^\prime)}a^{m-1}_k\ket{k}      \nonumber \\
   & & +e^{-\mathrm{i}(\theta_ml_m T_m+E_{m+1}1\tau_m^\prime)}\mathrm{i}  \nonumber \\
   & &   \times \sin(\Omega_m l_m\Delta_m)a^{m-1}_1\ket{m+1}\nonumber \\
   &\equiv& \sum^{m+1}_{k=1}a^m_k\ket{k},
\end{eqnarray}
from which we find the recursion relations
\begin{eqnarray}\label{recursion-m-1-m-1}
   && a^m_1= e^{-\mathrm{i}(\theta_ml_m T_m+E_1\tau'_m)}\cos(\Omega_m l_m\Delta_m)a^{m-1}_1,
             \label{recursion-m-1-m-1} \\
   && a^m_k =e^{-\mathrm{i}E_k( l_m T_m+\tau'_m)}a^{m-1}_k,   \qquad (2 \leq k\leq m),
             \label{recursion-m-1-m-2}   \\
   && a^m_{m+1}=\mathrm{i}e^{-\mathrm{i}(\theta_ml_m T_m+E_{m+1}\tau'_m)}
      \sin(\Omega_m l_m\Delta_m)a^{m-1}_1. \ \
             \label{recursion-m-1-m-3}
\end{eqnarray}

\subsection{Target state}

From (\ref{recursion-m-1-m-1}) and  (\ref{recursion-m-1-m-3}),
as well as (\ref{1-cycle-coefs}), we can easily obtain that
\begin{eqnarray}
  &&a^m_1 = \exp\left[-\mathrm{i}\sum^m_{i=1}(\theta_il_i T_i+E_1\tau'_i)\right]
           \prod^m_{i=1}\cos(\Omega_il_i\Delta_i),               \label{v-1}  \\
  &&a^m_2 = \exp\left\{-\im \left[E_2\sum^m_{i=2}(l_i T_i+\tau'_i)+\theta_1l_1\tau_1+E_2\tau'_1\right]
           \right\} \nonumber \\
  && \quad\quad\quad \times \mathrm{i}\sin(\Omega_1l_1\Delta_1),  \label{v-2}\\
  && a^m_{m+1}=\exp\left[-\mathrm{i}\left(\sum^m_{i=1}\theta_il_i T_i
     +\sum^{m-1}_{i=1}E_1\tau'_i+E_{m+1}\tau'_m\right)\right]                \nonumber  \\
  &&\quad\quad\quad\times \mathrm{i}\sin(\Omega_ml_m\Delta_m)\prod^{m-1}_{i=1}
  \cos(\Omega_il_i\Delta_i). \label{v-3}
\end{eqnarray}
Using (\ref{recursion-m-1-m-2}), we obtain
\begin{eqnarray}\label{m-cycle-k-coef}
   a^m_k &=& e^{-\mathrm{i}E_k (l_m T_m+\tau'_m)}\cdots e^{-\mathrm{i}E_k (l_k T_k+\tau'_k)} a^{k-1}_k
             \nonumber \\
    &=& \exp\left\{-\mathrm{i}\left[E_k \sum^m_{i=k-1}\tau'_i+
        E_1\sum^{k-2}_{i=1}\tau'_i+E_k\sum^m_{i=k}l_i T_i \right. \right. \nonumber \\
    & & + \left. \left. \sum^{k-1}_{i=1}\theta_il_i T_i\right]\right\}
        \mathrm{i}\sin(\Omega_{k-1}l_{k-1}\Delta_{k-1})    \nonumber   \\
    & & \times\prod^{k-2}_{i=1}\cos(\Omega_il_i\Delta_i),
        \quad 3\leq k \leq m,
\end{eqnarray}
where we have used (\ref{v-3}) with $m$ replaced by $k-1$.
%{\color{red} We remark that, if replace $k$ with $m+1$, one will find that
%(\ref{m-cycle-k-coef}) include the case $k=m+1$, namely, $3\leq k\leq m+1$.}

After $N-1$ cycles, or letting $m=N-1$ in elements (\ref{v-1}-\ref{v-3}) and
(\ref{m-cycle-k-coef}), we find the target states
\begin{equation}\label{N-1-cyc-final-sta}
  \ket{\psi^\prime}^{N-1}=\sum^N_{k=1}a^{N-1}_k\ket{k}=\sum^N_{k=1}C_k \gamma_k \ket{k},
\end{equation}
where the {\em real} probability amplitudes $C_k$ are
\begin{eqnarray}\label{realProbabilityAmplitude}
     C_1&=&\prod^{N-1}_{i=1}\cos(\Omega_il_i\Delta_i), \nonumber \\
     C_2&=&\sin(\Omega_1l_1\Delta_1) , \nonumber\\
     C_k&=&\sin(\Omega_{k-1}l_{k-1}\Delta_{k-1})\prod^{k-2}_{i=1}\cos(\Omega_il_i\Delta_i), \nonumber \\
        & & (3\leq k\leq N),
\end{eqnarray}
and phases $\gamma_k$ are
\begin{eqnarray}\label{relativePhase}
   \gamma_1 &=& \exp\left[-\mathrm{i}\sum^{N-1}_{i=1}(\theta_il_i T_i+E_1\tau'_i)\right] ,  \nonumber \\
   \gamma_2 &=& \mathrm{i}\exp\left\{-\mathrm{i}\left[E_2\sum^{N-1}_{i=2}(l_i T_i+\tau'_i)+
                \theta_1 l_1 T_1+E_2\tau'_1\right]\right\},\nonumber  \\
   \gamma_k &=& \exp\left\{-\mathrm{i}\left[E_k \sum^{N-1}_{i=k-1}\tau'_i+E_1\sum^{k-2}_{i=1}\tau'_i
                +E_k\sum^{N-1}_{i=k}l_i T_i   \right.\right.  \nonumber \\
            & & \left.\left. + \sum^{k-1}_{i=1}\theta_il_i T_i\right]\right\}\mathrm{i},
                \qquad (3\leq k\leq N).
\end{eqnarray}

%where the probability amplitudes are obtained as
%\begin{eqnarray}\label{N-system-coeffic}
%   a_1^{N-1}&=& \exp\left[-\mathrm{i}\sum^{N-1}_{i=1}(\theta_il_i T_i+E_1\tau'_i)\right]
%                \prod^{N-1}_{i=1}\cos(\Omega_il_i\Delta_i), \nonumber\\
%   a_2^{N-1}&=& \exp\left\{-\mathrm{i}\left[E_2\sum^{N-1}_{i=2}(l_iT_i+\tau'_i)+
%                \theta_1l_1 T_1+E_2\tau'_1\right]\right\}\nonumber \\
%            & & \times \mathrm{i}\sin(\Omega_1l_1\Delta_1),                  \nonumber\\
%   a_k^{N-1}&=& \exp\left\{-\mathrm{i}
%          \left[E_k \sum^{N-1}_{i=k-1}\tau'_i+E_1\sum^{k-2}_{i=1}\tau'_i \right.\right.\nonumber\\
%          && \left.\left.+ E_k\sum^{N-1}_{i=k}l_i T_i+ \sum^{k-1}_{i=1}\theta_il_i T_i\right]\right\}                       \nonumber   \\
%      & & \times \mathrm{i}\sin(\Omega_{k-1}l_{k-1}\Delta_{k-1})\prod^{k-2}_{i=1}\cos(\Omega_il_i\Delta_i),
%          \nonumber\\
%      &&  (3\leq k\leq N).
%\end{eqnarray}

\subsection{Determine control parameters}

% To determine control parameters $l_i\Delta_i,\,\tau'_i, \,1\leq i\leq N-1$, %we write
%$a_n = \gamma_n C_n$, where $\gamma_n$ are phases and $C_n$ the real probability amplitude,
%\begin{equation}\label{stateSuperposition}
%     \ket{\psi}=\sum^N_{n=1}a_n \ket{n}=\sum^N_{n=1}\gamma_n C_n \ket{n},
%\end{equation}
% namely
%\begin{eqnarray}\label{realProbabilityAmplitude}
%     C_1&=&\prod^{N-1}_{i=1}\cos(\Omega_il_i\Delta_i) \nonumber \\
%     C_2&=&\sin(\Omega_1l_1\Delta_1) , \nonumber\\
%     C_k&=&\sin(\Omega_{k-1}l_{k-1}\Delta_{k-1})\prod^{k-2}_{i=1}\cos(\Omega_il_i\Delta_i), \nonumber \\
%        & & (3\leq k\leq N).
%\end{eqnarray}
%and phase $\gamma_n$
%\begin{eqnarray}\label{relativePhase}
%   \gamma_1 &=& \exp\left[-\mathrm{i}\sum^{N-1}_{i=1}(\theta_il_i T_i+E_1\tau'_i)\right] ,  \nonumber \\
%   \gamma_2 &=& \mathrm{i}\exp\left\{-\mathrm{i}\left[E_2\sum^{N-1}_{i=2}(l_i T_i+\tau'_i)+
%                \theta_1 l_1 T_1+E_2\tau'_1\right]\right\},\nonumber  \\
%   \gamma_k &=& \exp\left\{-\mathrm{i}\left[E_k \sum^{N-1}_{i=k-1}\tau'_i+E_1\sum^{k-2}_{i=1}\tau'_i
%                +E_k\sum^{N-1}_{i=k}l_i T_i   \right.\right.  \nonumber \\
%            & & \left.\left. + \sum^{k-1}_{i=1}\theta_il_i T_i\right]\right\}\mathrm{i},
%                \qquad (3\leq k\leq N)
%\end{eqnarray}

For a given target state, namely, $C_n$ and $\gamma_n$ are given, we can determine
the control parameters $\{l_n\Delta_n,\tau'_n|n=1,2,...,N-1  \}$. From $C_2$ we can
determine $l_1\Delta_1$ and then $l_2\Delta_2$ from $C_3$, until all $l_{k-1}\Delta_{k-1}$
obtained recursively.

As for $\tau^\prime_i$, we first obtain $\sum_{i=1}^{N-1} \tau_i^\prime$ from $\gamma_1$ or
$\gamma_2$ and $E_3 \sum_{i=2}^{N-1}\tau^\prime_i+E_1 \tau_1^\prime$ from $\gamma_3^\prime$.
As $E_1\neq E_3$, we can obtain $\tau^\prime_1$ and $\sum_{i=2}^{N-1}\tau^\prime_i$. From
$\gamma_4$, we obtain  $E_4 \sum_{i=3}^{N-1}\tau^\prime_i+E_1 \tau_2^\prime$ from which
we obtain $\tau^\prime_2$ and $\sum_{i=3}^{N-1}\tau^\prime_i$. Recursively, we can obtain
all $\tau^\prime_i$'s.

\section{Control of System II} \label{section-4dimensional}

We now turn to control of system II. In the first cycle, we apply the field
with frequency $\omega_1 = E_2-E_1=E_4-E_3$ to interact
with the system. The total Hamiltonian corresponding to positive and negative pulses reads
\begin{eqnarray}
&& H^\pm=H_1^\pm +H_2^\pm, \\
&& H_1^\pm= E_1 \ketbra{1}{1} +E_2\ketbra{2}{2} \pm d_1 \left(\ketbra{1}{2}+\ketbra{2}{1}\right),\\
&& H_2^\pm= E_3 \ketbra{3}{3} +E_4\ketbra{4}{4} \pm d_2 \left(\ketbra{3}{4}+\ketbra{4}{3}\right),
\end{eqnarray}
satisfying $[H_1^\pm, H_2^\pm]=0$. Therefore the time evolution operator can be factorized
as
\begin{equation}
U_\pm(t)=U_\pm^1(t)U_\pm^2(t)=\exp\left(-\im H^\pm_1t\right)\exp\left(-\im H^\pm_2t\right).
\end{equation}
Both operators $U_\pm^1(t)$ and $U_\pm^2(t)$ can be treated similarly as in the
Sec.~\ref{sec-pulse-evolution-operator}.
But if we prepare the system initially on the state $\ket{1}$, we have
$H_2^\pm \ket{1}=0$ and thus $U_\pm ^2 (t)=\ket{1}$. In
this case
\begin{equation}
U_\pm(t)\ket{1}=U_\pm^1(t) \ket{1}=\exp\left(-\im H^\pm_1t\right) \ket{1}.
\end{equation}
To expand $U^\pm_1(t)$, we rewrite $H_1^\pm$ as
\begin{eqnarray}
&& H_1^\pm =
\theta_{12}I - \theta_{12} I_{34}
     +H^{(12)}_{c\pm},
\end{eqnarray}
where $I_{34}=\ketbra{3}{3}+\ketbra{4}{4}$, and
\[
 H^{(12)}_{c\pm} \equiv \frac{1}{2} \omega_1 \left(\ketbra{2}{2}-\ketbra{1}{1}\right)
\pm d_1\left(\ketbra{1}{2}+\ketbra{2}{1}\right).
\]
Then we have
\begin{eqnarray}
&& e^{-\im H_1^\pm t}
   = \left[ I + \left[\cos(\Omega_1t)-1\right]I_{12}
     -\frac{\im}{\Omega_1}\sin(\Omega_1t)H^{(12)}_{c\pm} \right] \nonumber \\
&& \qquad     \times \exp{\left[-\im \theta_{12} I \right]}
  \exp{\left[ \im \theta_{12}(\ketbra{3}{3}+\ketbra{4}{4})\right]},
\end{eqnarray}
where $\Omega_1 = \left[d_1^2 + \omega_1^2/4 \right]^{1/2}$.
Acting on the initial state $\ket{1}$, we have
\begin{equation}
e^{-\im H_1^\pm t}\ket{1}
=e^{\theta_{12}t}\left[\cos(\Omega_1t)\ket{1}
\mp \im \sin(\Omega_1t)\ket{2}\right],
\end{equation}
where we have used the strong field approximation, namely
$d_1\gg\omega_1$, $\omega_1/\Omega_1\rightarrow 0$, $d_1/\Omega_1\rightarrow 1$.

We control the system for a time period $l_1 T_{1}$, yielding
\begin{eqnarray}\label{U-effect-on-state1}
\ket{\psi}^{(1)}
&=& U^{(1)}(l_1T_1)\ket{1}
    =\left(U^{-}_1 (\tau^1_2) U^{+}_1 (\tau^1_1)\right)^{l_1}\ket{1}\nonumber   \\
&=& e^{-\im\theta_{12}l_1T_1}\left[\cos(\Omega_1l_1\Delta_1)\ket{1}+\im\sin(\Omega_1l_1\Delta_1)\ket{2}\right],
\nonumber
\end{eqnarray}
with $\Delta_1=\tau^1_2-\tau^1_1$. After free evolution for $\tau'_1$, the state of system is
\begin{eqnarray} \label{1-cyc-free-evol}
\ket{\psi'}^{(1)} = e^{-\mathrm{i}H_0\tau'_1}\ket{\psi}^{(1)}
     =a^1_1\ket{1}+a^1_2\ket{2},
\end{eqnarray}
where
\begin{eqnarray}\label{1-cycle-coef}
   && a^1_1=e^{-\mathrm{i}(\theta_{12}l_1T_1+E_1\tau'_1)}\cos(\Omega_1 l_1\Delta_1) ,       \nonumber   \\
   && a^1_2=e^{-\mathrm{i}(\theta_{12}l_1T_1+E_2\tau'_1)}\mathrm{i}\sin(\Omega_1 l_1\Delta_1).
\end{eqnarray}

In cycle 2, we apply the field with frequency $\omega_2 = E_4-E_1$ to control
the system for a time period $l_2 T_{2}$. Standard treatment for the time evolution operator
in Sec.~\ref{sec-pulse-evolution-operator} applies
in this case and we can easily find the state of the system after interaction
with control field and free evolution for time period $\tau'_2$
\begin{eqnarray} \label{2-cyc-free-evol}
\ket{\psi'}^{(2)} & \equiv & e^{-\mathrm{i}H_0\tau'_2}\left(U^{(2)}_-(\tau^2_2)U^{(2)}_+(\tau^2_1)\right)^{l_2}\ket{\psi'}^{(1)}    \nonumber\\
    &=&  a^2_1\ket{1}+a^2_2\ket{2}+a^2_4\ket{4},
\end{eqnarray}
where
\begin{eqnarray}\label{recursion-cyc-2}
a^2_1 &=&e^{-\mathrm{i}(\theta_{14}l_2T_2+\theta_{12}l_1T_1+E_1\sum^2_{i=1}\tau'_i)} \nonumber\\
      & & \times\cos(\Omega_2 l_2\Delta_2)\cos(\Omega_1 l_1\Delta_1) ,          \nonumber\\
a^2_2 &=& e^{-\mathrm{i}(\theta_{12}l_1T_1+E_2l_2T_2+E_2\sum^2_{i=1}\tau'_i)}
          \mathrm{i}\sin(\Omega_1 l_1\Delta_1)  ,                            \nonumber\\
a^2_4 &=& e^{-\mathrm{i}(\theta_{14}l_2T_2+\theta_{12}l_1T_1+E_4\tau'_2+E_1\tau'_1)} \nonumber\\
   && \times \mathrm{i}\sin(\Omega_2 l_2\Delta_2)\cos(\Omega_1 l_1\Delta_1).
\end{eqnarray}

In cycle 3, we apply the field with frequency $\omega_3 = E_3-E_2$ to control the system
for a period $l_3 T_{3}$ and then allow the system for a free evolution for time
period $\tau_3^\prime$. The time evolution operator can be treated as in Sec.~\ref{sec-pulse-evolution-operator} and the
target state can be obtained as
\begin{eqnarray}\label{3-cycle}
\ket{\psi^\prime}^{(3)}&=&e^{-\mathrm{i}H_0\tau'_3}\left(U^{(3)}_-(\tau^3_2)U^{(3)}_+(\tau^3_1)\right)^{l_3}\ket{\psi'}^{(2)} \nonumber \\
    &=& \sum^4_{n=1}a_n \ket{n}=\sum^4_{n=1}\gamma_n C_n \ket{n},
\end{eqnarray}
where the real probability amplitudes are
\begin{eqnarray}\label{realProbabilityAmplitude}
     C_1&=&\cos(\Omega_2 l_2\Delta_2)\cos(\Omega_1 l_1\Delta_1),  \nonumber \\
     C_2&=&\cos(\Omega_3 l_3\Delta_3)\sin(\Omega_1 l_1\Delta_1),  \nonumber\\
     C_3&=&\sin(\Omega_3 l_3\Delta_3)\sin(\Omega_1 l_1\Delta_1), \nonumber\\
     C_4&=&\sin(\Omega_2 l_2\Delta_2))\cos(\Omega_1 l_1\Delta_1),
\end{eqnarray}
and the relative phases are
\begin{eqnarray}\label{relativePhase}
   \gamma_1 &=& \exp\left[-\mathrm{i}\left(\theta_{14}l_2T_2+\theta_{12}l_1T_1+
                E_1(\tau'_1+\tau'_2+\tau'_3)\right.\right. \nonumber \\
            & & \left.\left. +E_1l_3T_3 \right)\right], \\
   \gamma_2 &=& \mathrm{i}\exp\left[-\mathrm{i}\left(\theta_{23}l_3T_3
                +\theta_{12}l_1T_1+E_2l_2T_2 \right.\right. \nonumber \\
            & & \left.\left. +E_2(\tau'_1+\tau'_2+\tau'_3)\right)\right],\\
   \gamma_3 &=& \exp\left[-\mathrm{i}\left(\theta_{23}l_3T_3
                +\theta_{12}l_1T_1+E_3\tau'_3 \right.\right. \nonumber \\
            & & \left.\left. +E_2 (\tau'_1+\tau_2^\prime) +E_2l_2T_2\right)\right],  \\
   \gamma_4 &=& \mathrm{i}\exp\left[-\mathrm{i}\left(\theta_{14}l_2T_2
                +\theta_{12}l_1T_1 \right.\right. \nonumber\\
            & & \left.\left.  +E_4 (\tau'_2+\tau_3^\prime) +E_4l_3T_3+E_1\tau'_1 \right)\right].
\end{eqnarray}

We need to determine the control parameters $\{l_n\Delta_n, \tau'_n \ |\ n=1,2,3\}$ from the
probability amplitudes of the target state. It is easy to find that
\begin{eqnarray}
\frac{C_4}{C_1} = \tan(\Omega_2 \Delta_2 l_2), \qquad
\frac{C_3}{C_2} = \tan(\Omega_3 \Delta_3 l_3),
\end{eqnarray}
from which we can find control parameters $\Delta_2 l_2$ and $\Delta_3 l_3$
and then determine $\Delta_1 l_1$ from $C_1$.

From $\gamma_1$, $\gamma_3$ and $\gamma_4$, we can determine
\begin{eqnarray}
&& \tau'_1+\tau'_2+\tau'_3, \nonumber \\
&& E_2(\tau'_1+\tau'_2)+E_3\tau_3^\prime, \nonumber \\
&& E_1 \tau_1^\prime + E_2 (\tau'_2+\tau'_3),
\end{eqnarray}
whose determinant of the coefficient matrix is $(E_2-E_1)(E_2-E_3)\neq 0$. So we
can find control parameters $\tau^\prime_i \ (i=1,2,3)$.

\section{System with equal energy gaps}

In this section we consider the simplest system with all equal energy gaps, the
3-dimensional system with $E_3-E_2=E_2-E_1=\mu$. To control this system, we
apply pulse field with frequency $\omega=\mu$ to drive the system for time
period $l_1 T_1$ and then let it involve for a time period $\tau_1^\prime$,
in the first cycle. In cycle 2, we use control field with frequency
$\omega_2 = E_3-E_1 = 2\mu$ for time period $l_2 T_2$ first and then
leave it for a free evolution for time period $\tau^\prime_2$.

\subsection{Cycle 1}

The Hamiltonians between the system and control field corresponding to
positive and negative pulses are
\begin{eqnarray}
H^{(1)}_\pm &=&\sum_{i=1}^3 E_i \ketbra{i}{i}\pm \left[ d_1(\ketbra{1}{2}+\ketbra{2}{1})\right. \nonumber \\
&& \left. +d_2(\ketbra{2}{3}+
\ketbra{3}{2})\right].
\end{eqnarray}
The last two terms are not commutative each other and thus the corresponding time evolution
operator cannot be treated as in last section. We note that $\im H^{(1)}_\pm$ is an
element of u(3) Lie algebra,
and its time evolution operator can be generally written as the product of single-parameter
subgroup elements $\exp(-ig_i(t) x_i)$, where $x_i$ are all basis elements of u(3), according to
Wei-Norman theorem \cite{wei}. One needs to solve a set of nonlinear equations
to determine parameter $g_i(t)$, which is generally difficult.

However, when the system is initially prepared on the ground state
$\ket{1}$, it is enough to find $U^\pm(t)\ket{1}$. It is easy to find that
\begin{eqnarray}
&& H^{(1)}_\pm \ket{1}=E_1\ket{1}+d_1\ket{2}, \nonumber \\
&& (H^{(1)}_\pm)^2 \ket{1}= \left(E_1^2+d_1^2\right)\ket{1}+d_1(E_1+E_2)\ket{2}+d_1d_2\ket{3}. \nonumber
\end{eqnarray}
Suppose that $E_2=0$, without losing generality, then $E_1=-\mu, E_3=\mu \ll d_1,d_2$,
under strong field approximation. Then we have
\begin{eqnarray}
&& H^{(1)}_\pm \ket{1}= d_1\ket{2}, \nonumber \\
&& \left(H^{(1)}_\pm \right)^2 \ket{1}=  d_1^2 \ket{1} +d_1 d_2\ket{3}.
\end{eqnarray}
It is not difficult to find that
\begin{eqnarray}
&& (H^{(1)}_\pm)^{2k+1} \ket{1} = \pm d_1   \Omega_1^{2k} \ket{2}, \quad k\geq 0, \\
&& \left(H^{(1)}_\pm \right)^{2k} \ket{1} = d_1  \Omega_1^{2(k-1)} (d_1\ket{1}
    + d_2 \ket{3}), \ \ k>0, \ \
\end{eqnarray}
where $\Omega_1 = \sqrt{d_1^2 + d_2^2}$. Therefore, the time evolution
operator acting on the initial state $\ket{1}$ is obtained as
\begin{eqnarray}
e^{-\im H^{(1)}_\pm t}\ket{1}
&=& \left\{\frac{d_1}{\Omega_1^2} \left[\cos\left(\Omega_1 t\right)-1\right]+1\right\}\left(d_1\ket{1}+d_2\ket{3}\right)
    \nonumber \\
& & \mp \im \frac{d_1}{\Omega_1} \sin(\Omega_1 t) \ket{2}.
\end{eqnarray}
Similarly, we can further obtain that
\begin{eqnarray}
&& (H^{(1)}_\pm)^{(2k+1)} \ket{2} = \pm \Omega_1^{2k}(d_1\ket{1}+d_2\ket{3}), \nonumber\\
&& (H^{(1)}_\pm)^{2k}\ket{2}=\Omega_1^{2k}\ket{2},
\end{eqnarray}
and
\begin{eqnarray}
e^{-\im H^{(1)}_\pm t}\ket{2}
&=& \cos\left(\Omega_1 t\right)\ket{2} \nonumber \\
& &    \mp \im \frac{\sin(\Omega_1 t)}{\Omega_1} (d_1\ket{1} + d_2\ket{3}).
\end{eqnarray}
We also have
\begin{eqnarray}
e^{-i H^{(1)}_\pm t}\ket{3}
&=& \frac{d_2}{\Omega_1^2} \left[\cos\left(\Omega_1 t\right)-1\right](d_1\ket{1}+d_2\ket{3})
    \nonumber \\
& & \mp i \frac{d_2}{\Omega_1} \sin(\Omega_1 t) \ket{2}+ \ket{3}.
\end{eqnarray}
Then we obtain the action of the time evolution operator in a period $T_1$
\begin{eqnarray}
   &&e^{ -\im H^{(1)}_-\Delta_1^2 }e^{ -\im H^{(1)}_+\Delta_1^1 }\ket{1} \nonumber\\
   && \quad =\left\{\frac{d^2_1}{\Omega_1^2}[\cos(\Omega_1\Delta_1)-1]+1\right\}\ket{1}+\frac{\mathrm{i}d_1}{\Omega_1}\sin(\Omega_1\Delta_1)\ket{2} \nonumber\\
   && \quad\quad +\frac{d_1d_2}{\Omega_1^2}[\cos(\Omega_1\Delta_1)-1]\ket{3},
\end{eqnarray}
where $\Delta_1=\Delta_1^2 -\Delta_1^1 $.
After the second pulse, we have
\begin{eqnarray}
   &&\left(e^{ -\im H^{(1)}_-\Delta_1^2 }e^{ -iH^{(1)}_+\Delta_1^1 }\right)^2\ket{1} \nonumber\\
   &&=\left\{\frac{d^2_1}{\Omega_1^2}[\cos(\Omega_12\Delta_1)-1]+1\right\}\ket{1}+\frac{\mathrm{i}d_1}{\Omega_1}\sin(\Omega_12\Delta_1)\ket{2} \nonumber\\
   && \quad +\frac{d_1d_2}{\Omega_1^2}[\cos(\Omega_12\Delta_1)-1]\ket{3}.
\end{eqnarray}
Recursively, when $l_1=\tau_1/T_1$ pulses are applied, the system is driven to
\begin{eqnarray}
   &&\ket{\psi}^{(1)}=\left(e^{ -\im H^{(1)}_-\Delta_1^2 }e^{ -\im H^{(1)}_+\Delta_1^1 }\right)^{l_1}\ket{1} \nonumber\\
   &&=\left\{\frac{d^2_1}{\Omega_1^2}\left[\cos(\Omega_1l_1\Delta_1)-1\right]+1\right\}\ket{1}+\frac{\mathrm{i}d_1}{\Omega_1}\sin(\Omega_1l_1\Delta_1)\ket{2} \nonumber\\
   && \quad +\frac{d_1d_2}{\Omega_1^2}\left[\cos(\Omega_1l_1\Delta_1)-1\right]\ket{3}.
\end{eqnarray}
After free evaluation for $\tau'_1$, the system is driven to
\begin{eqnarray}
   \ket{\psi'}^{(1)}&=a^1_1\ket{1}+a^1_2\ket{2}+a^1_3\ket{3},
\end{eqnarray}
where
\begin{eqnarray}\label{recursion-cyc-2}
  a^1_1&=& e^{-\mathrm{i}E_1\tau'_1}\re (a_1^1),\nonumber\\
   a^1_2 &=&e^{-\mathrm{i}E_2\tau'_1}\frac{\mathrm{i}d_1}{\Omega_1}\sin(\Omega_1l_1\Delta_1) ,       \nonumber\\
   a^1_3&=&e^{-\mathrm{i}E_3\tau'_1} \re(a^1_3),
\end{eqnarray}
and
\begin{eqnarray}
   \re(a_1^1) &\equiv & \frac{d^2_1}{\Omega_1^2}\left[\cos(\Omega_1l_1\Delta_1)-1\right]+1 ,\\
   \re(a^1_3) &\equiv & \frac{d_1d_2}{\Omega_1^2}\left[\cos(\Omega_1l_1\Delta_1)-1\right].
\end{eqnarray}

\subsection{Cycle 2}
In cycle 2, the total Hamiltonian is
 \begin{eqnarray}
   & H^{(2)}_\pm=\sum^3_{i=1}E_i\ketbra{i}{i}\pm d_3(\ketbra{3}{1}+\ketbra{1}{3}),
\end{eqnarray}
which causes transition between states $\ket{1}$ and $\ket{3}$. So when $l_2=\tau_2/T_2$
pulses are applied to the system, we can use the result in
Sec.~\ref{sec-pulse-evolution-operator} to obtain that
\begin{eqnarray}
&& U^{(2)}(l_2T_2)\ket{1}=(U^{(2)}(T_2))^{l_2}\ket{1}\nonumber\\
&& \quad  =e^{-\mathrm{i}\theta_{13}l_2T_2}
        \left[\cos(\Omega_2l_2\Delta_2)\ket{1}+\mathrm{i}\sin(\Omega_2l_2\Delta_2)\ket{3}\right],\quad \nonumber \\
%%%%%%%%%%
&& U^{(2)}(l_2T_2)\ket{3}=(U^{(2)}(T_2))^{l_2}\ket{3}\nonumber\\
&& \quad =e^{-\mathrm{i}\theta_{13}l_2T_2}\left[\cos(\Omega_2l_2\Delta_2)\ket{3}
       +\mathrm{i}\sin(\Omega_2l_2\Delta_2)\ket{1}\right],\nonumber \\
&& U^{(2)}(l_2T_2)\ket{2}=\ket{2},
\end{eqnarray}
where $\Omega_2=\sqrt{\omega_2^2 /4 +d_3^2} $.
After the system interacts with $l_2$ pulses and then involves for time period $\tau_2^\prime$,
we arrive at the target state
%\begin{eqnarray}\label{t2-cycle}
%\ket{\psi}^{(2)} &=& \left(U^{(2)}(T_2)\right)^{l_2}\ket{\psi'}^{(1)}                                 \nonumber\\
%%   &=&a^1_1e^{-\mathrm{i}\theta_{13}l_2T_2}\left[\cos(\Omega_2l_2\Delta_2)\ket{1}+\mathrm{i}\sin(\Omega_2l_2\Delta_2)\ket{3}\right] \nonumber\\
%%   &&+a^1_3e^{-\mathrm{i}\theta_{13}l_2T_2}\left[\cos(\Omega_2l_2\Delta_2)\ket{3}+\mathrm{i}\sin(\Omega_2l_2\Delta_2)\ket{1}\right]    \nonumber\\
%%   &&+a^1_2e^{-\mathrm{i}E_2l_2T_2}\ket{2}\nonumber\\
%&=& e^{-\mathrm{i}\theta_{13}l_2T_2}\left[a^1_1\cos(\Omega_2l_2\Delta_2)+a^1_3\mathrm{i}\sin(\Omega_2l_2\Delta_2)\right]\ket{1} \nonumber\\
%& & +e^{-\mathrm{i}\theta_{13}l_2T_2}\left[a^1_1\mathrm{i}\sin(\Omega_2l_2\Delta_2)\right.\nonumber\\
%& & \left. +a^1_3\cos(\Omega_2l_2\Delta_2)\right]\ket{3} +a^1_2e^{-\mathrm{i}E_2l_2T_2}\ket{2}.
%\end{eqnarray}
%free evolution for $\tau'_2$, we arrive at the target state
\begin{eqnarray} \label{t2-cyc-free-evol}
\ket{\psi'}^{(2)}&=& e^{-\mathrm{i}H_0\tau'_2}\ket{\psi}^{(2)}
   = \sum_{k=1}^3 a^2_k \ket{k},
\end{eqnarray}
where
\begin{eqnarray}\label{t2-cyc-2}
%   a^2_1&=&e^{-\mathrm{i}(E_1\tau'_2+\theta_{13}l_2T_2)} \nonumber\\
%        & &\times \left[a^1_1\cos(\Omega_2l_2\Delta_2)+a^1_3\mathrm{i}\sin(\Omega_2l_2\Delta_2)\right], \\
   a_1^2 &=& e^{-\im E_1(\tau_1^\prime +\tau_2^\prime)}\re (a^1_1) \cos(\Omega_2l_2\Delta_2)\nonumber \\
      & &   + e^{-\im (E_3\tau_1^\prime + E_1 \tau_2^\prime} \im \re (a^1_3) \sin(\Omega_2l_2\Delta_2), \\
   a^2_2 &=&e^{-\mathrm{i}E_2(\tau'_1+\tau'_2)}\frac{\mathrm{i}d_1}{\Omega_1}\sin(\Omega_1l_1\Delta_1),\\
%%
%   a^2_3&=&e^{-\mathrm{i}(E_3\tau'_2+\theta_{13}l_2T_2)} \nonumber\\
%        & &\times\left[a^1_1\mathrm{i}\sin(\Omega_2l_2\Delta_2)+a^1_3\cos(\Omega_2l_2\Delta_2)\right].
   a_3^2 &=&  e^{-\im (E_1 \tau_1^\prime +E_3\tau_2^\prime)}\im \re (a^1_1) \sin(\Omega_2l_2\Delta_2)\nonumber \\
      & &   + e^{-\im E_3 (\tau_1^\prime + \tau_2^\prime)} \re (a^1_3) \cos(\Omega_2l_2\Delta_2),
\end{eqnarray}
where we have used the fact $\theta_{13}=0$.

\subsection{Determine control parameters}

From $a_2^2$, we can determine the control parameters $l_1\Delta_1$ and $\tau_1^\prime+\tau_2^\prime$,
as well as $\re(a_1^1), \re(a_3^1)$.
%Write
%\begin{eqnarray}
%a_1^2 &=& e^{-\im E_1(\tau_1^\prime +\tau_2^\prime)}\re (a^1_1) \cos(\Omega_2l_2\Delta_2)\nonumber \\
%      & &   + e^{-\im (E_3\tau_2^\prime + E_1 \tau_1^\prime} \im \re (a^1_3) \sin(\Omega_2l_2\Delta_2)\nonumber \\
%a_3^2 &=&  e^{-\im (E_1 \tau_1^\prime +E_3\tau_2^\prime)}\im \re (a^1_1) \sin(\Omega_2l_2\Delta_2)\nonumber \\
%      & &   + e^{-\im E_3 (\tau_1^\prime + \tau_2^\prime)} \re (a^1_3) \cos(\Omega_2l_2\Delta_2).
%\end{eqnarray}
Then one can check that
\begin{eqnarray}
&&-\im
\left[ e^{ \im E_1(\tau_1^\prime +\tau_2^\prime)}a_1^2 \re(a_3^1) -
e^{ -\im E_3(\tau_1^\prime +\tau_2^\prime)}(a_3^2)^* \re(a_1^1)\right] \nonumber \\
&& \quad = \left[
%e^{ -\im (E_3-E_1) (\tau_1^\prime+ \tau_2^\prime)}
(\re a^1_3)^2
+(\re a_1^1)^2
\right] \nonumber \\
&& \qquad \times e^{ -\im (E_3-E_1)\tau_1^\prime} \sin(\Omega_2l_2\Delta_2),
\end{eqnarray}
in which
both terms in middle brackets on the left and right hand sides are known for a given target state. So we can determine
$ e^{ -\im (E_3-E_1)\tau_1^\prime} \sin(\Omega_2l_2\Delta_2)$ and thus $\tau_1^\prime$
and $l_2\Delta_2$, as well as $\tau_2^\prime$.

\section{Conclusion}

In this paper we proposed protocols to control finite
dimensional quantum systems using square pulses. Time evolution
operators are explicitly obtained under strong field approximation
and used to control three types of finite dimensional systems.
Relationship between control parameters and probability amplitudes
of the target states are established via trigonometrical functions.
The control parameters are time periods of interaction between
the controlled system and control pulses and free evolution
time periods of the system itself.

We would like to remark that:
(1). Control protocol using square pulses avoid using the
       rotating wave approximation as using the harmonic field;
(2). Interaction between the system and control field does not supply
       the relative phases and the relative phases are achieved
       by free evolution of the controlled system between cycles.

As further works, we would like to generalize the investigation presented
in this paper to the indirect control protocol of
finite quantum systems, and the control protocol of quantum systems in the
presence of environment.

\section*{Acknowledgement}

This work is supported by the National Science Foundation of China under grand
numbers 11075108 and 61374057.


\vspace{0.5cm}

\begin{thebibliography}{99}
\bibitem{1}
V. P. Belavkin, {Automatica and Remote Control} {\bf 44}, 178188 (1983);
G. M. Huang, T. J. Tarn and J. W. Clark, {J. Math. Phys.} {\bf 24}, 2608 (1983).

\bibitem{ref1}
{\it Information Complexity and Control in Quantum Physics}, edited
by A. Blaquiere, S. Dinerand, and G. Lochak (Springer,
New York, 1987);
A. G. Butkovskiy and Yu. I. Samoilenko, {\it Control of Quantum-
Mechanical Processes and Systems} (Kluwer Academic, Dordrecht,
1990);
V. Jurdjevic, {\it Geometric Control Theory} (Cambridge University
Press, Cambridge, U.K., 1997);
S. Lloyd, {\it Phys. Rev. A} {\bf 62}, 022108 (2000).


\bibitem{optimal}
  M. A. Daleh, A. M. Peirce, and H. Rabitz, {Phys. Rev. A} {\bf 37}, 4950 (1988);
  A. Bartana, R. Kosloff, and D. J. Tannor, {Chem. Phys.} {\bf 267}, 95 (2001);
  U. Boscain, G. Charlot, J.-P. Gauthier, S. Duerin and H.-R. Jauslin,
  {J. Math. Phys.} {\bf 43}, 2107 (2002).

\bibitem{controllability} V. Ramakrishna and H. Rabitz, {Phys. Rev. A} {\bf 54}, 1715 (1996).

\bibitem{fu1} S. G. Schirmer, H. Fu and A. I. Solomon, {Phys. Rev. A} {\bf 63}, 063410 (2001).

\bibitem{fu2} H. Fu, S. G. Schirmer and A. I. Solomon, {J. Phys. A: Math. Gen.} {\bf 34}, 1679 (2001).

\bibitem{graph} G. Turinici, Mathematical Models and Methods for ab Initio Quantum Chemistry,
(Lecture Notes in Chemistry, Vol. 74, 2000), ed. M. Defranceschi and C. Le Bris, Springer, Berlin;
G. Turinici and H. Rabitz, Chem. Phys. {\bf 267}, 1 (2001).

\bibitem{feedback} H. M. Wiseman and G. J. Milburn, {\it Quantum measurement and control}
(Cambridge University Press, Cambridge, U.K., 2010).

\bibitem{indirect1} H. C. Fu, H. Dong, X. F. Liu and C. P. Sun, {Phys. Rev. A.} {\bf 75}, 052317 (2007).
\bibitem{indirect2} H. C. Fu, H. Dong, X. F. Liu and C. P. Sun, {J. Phys. A. Math. Gen} {\bf 42}, 045303 (2009).

\bibitem{romano} R. Romano and D. D'Alessandro, {Phys. Rev. Lett.} {\bf 97}, 080402 (2006);
{Phys. Rev. A} {\bf 73}, 022323 (2006).

\bibitem{penchen} A. Pechen and H. Rabitz, {Phys. Rev. A} {\bf 73}, 062102 (2006).

\bibitem{cartan} R. Romano and D. D'Alessandro, {J. Phys. A.} {\bf 40}, 2439 (2007).

\bibitem{tang1} J. Tang and H. C. Fu, Commun. Theor. Phys. {\bf 60}, 731 (2013).

\bibitem{b} M. Scully and M.S. Zubairy, {\it Quantum Optics}
(Cambridge University Press, Cambridge, U.K., 2000).

\bibitem{wei} J. Wei and E. Norman, {J. Math. Phys.} {\bf 4}, 575 (1963).
\end{thebibliography}
\end{document}